# Using FLAME Toolkit for Agent-Based Simulation:

# Case Study Sugarscape Model

**Mariam Kiran**
**School of Electrical Engineering and Computer Science**
**University of Bradford**
**Richmond Road,**
**Bradford, BD7 4DP**
**m.kiran@bradford.ac.uk**

**Abstract:**
Social scientists have used agent-based models to understand how individuals interact and behave in various political, ecological and economic scenarios. Agent-based models are ideal for understanding such models involving interacting individuals producing emergent phenomenon. Sugarscape is one of the most famous examples of a social agent-based model which has been used to show how societies grow in the real world.

This paper builds on the Sugarscape model, using the Flexible Large scale Agent-based modelling Environment (FLAME) to simulate three different scenarios of the experiment, which are based on the Sugar and Citizen locations. FLAME is an agent-based modelling framework which has previously been used to model biological and economic models. The paper includes details on how the model was written and the various parameters set for the simulation. The results of the model simulated are processed for three scenarios and analysed to see what affect the initial starting states of the agents had on the overall result obtained through the model and the variance in simulation time of processing the model on multicore architectures.

The experiments highlight that there are limitations of the FLAME framework and writing simulation models in general which are highly dependent on initial starting states of a model, also raising further potential work which can be built into the Sugarscape model to study other interesting phenomenon in social and economic laws.

# 1 Introduction

Decentralised control is an important aspect of self-organising systems. For instance, insect colonies have been studied to deduce how, despite each insect working independently, the colony collectively works extremely efficiently [23]. Termites and ants are examples of such colonies which emerge into a complicated but precise execution action plan.

Growing artificial agent societies has been a useful technique to study how societies are created and how they thrive in changing real world conditions. Examples include growing cell tissues on petri dishes or hosting an ant colony in a tank. Computer simulations are an alternate method to study the behavior of organisms interacting together in a safe environment which can be validated with experimental data. These simulations have used parallel computing technologies to make more complicated models, particularly using agent-based modelling. One of the famous example of a social agent-based model was written by Epstein et al. [8]. This model used an extremely simple setting for a society of agents playing very simple rules but allowed very complex research to be conducted investigating societal issues. The experiment called the Sugarscape model, contained an artificial society of individual agents who were allowed to move around on a 2D grid space and look for sugar [13]. Figure 1 depicts screen shots of the model setting before the simulation begins. The sugar has been distributed in two piles on opposite corners of the scenario with the agents distributed randomly.

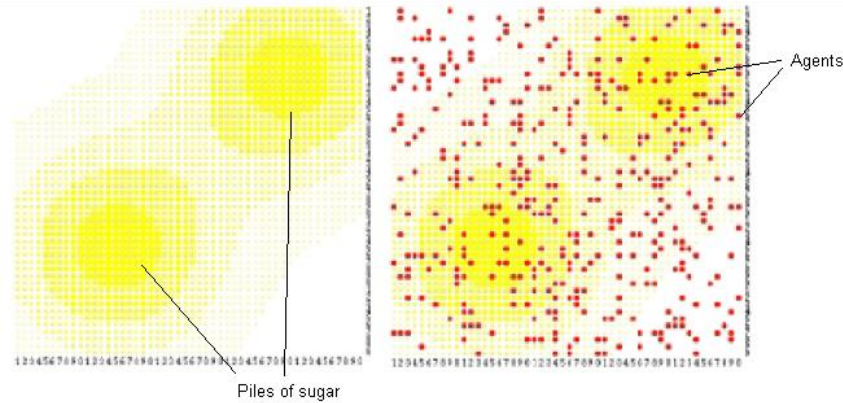

Fig. 1: Initial distribution in the Sugarscape model (left) and initial distribution with agents (right). Adapted from [3].

Each agent or citizen has an area of vision and moving distance for capturing the sugar where the agent would make decisions locally seeing its circle of influence or bounded view. This made the agent society in Sugarscape highly decentralised in the same way as ant or termite colonies are in nature. The model can then be analysed on a larger scale, where various patterns emerging from other societies can be observed. The traditional Sugarscape model allowed the agents to see in four directions to search for sugar (Figure 2). The agents are laid out on a grid structure represented using cellular automata, where using rules of only looking north, south, east and west, agents can move to a sugar laden square to eat it. Every individual is selfish and the information each individual has is different making decisions on what the individual knows. This gives rise to bounded rationality, where there is rationality depending on the bounds of the individual's information space [21].

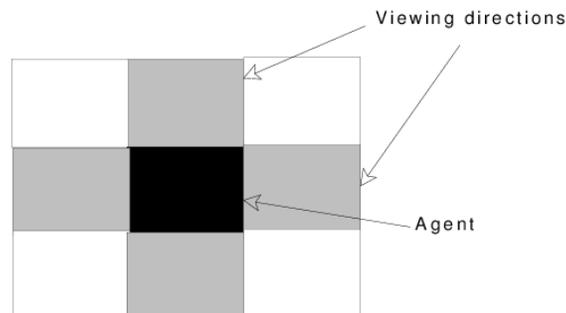

Fig. 2: Agent perception in traditional Sugarscape. The agent can see north, south, east and west only.

With the inclusion of location the Sugarscape model showed how the spatial distribution or landscape can influence the inhabitant agents and sugars. The agents were seen crawling over the landscape during the simulation looking for sugar as seen in Figure 3. The sugar acts as a source of energy which is distributed in two piles over the landscape. The agents also have a metabolism which uses up part of the sugar each time they move. The basic rules followed by the agents were:
1. Agent look north, south, east or west for sugar.
2. If sugar is found, move to the sugar location and eat.

3. If sugar is not found, randomly move to another square.

Each time the agent moved, a small amount of its stored sugar was used by metabolism. Eventually having used up all the sugar, the agents would die and disappear from the scenario.

This paper reconstructs the Sugarscape model in FLAME the large-scale parallelisation toolkit which allows agent-based models to automatically parallelise for faster simulations. The experiment is tested in various scenarios to test the influence it has on the simulation times of the model. This paper has been organised in the following manner: After the introduction, Section 2 talks in detail about the Sugarscape model and various research it has been used in.

Section 3 presents the experiment with a description of the FLAME framework and how it was used to model the Sugarscape model presenting the algorithm for modelling the agents in the model. The accompanying C code for the agents has been attached as appendix at the end of this paper. Section 4 presents the results and a discussion with an additional analysis on the simulation times of the three experiments. With the related work presented in Section 5, we finally conclude with Section 6 with future recommendations and limitations of the experiment. Further graphs are presented in the Appendix to help with analysis of the results.

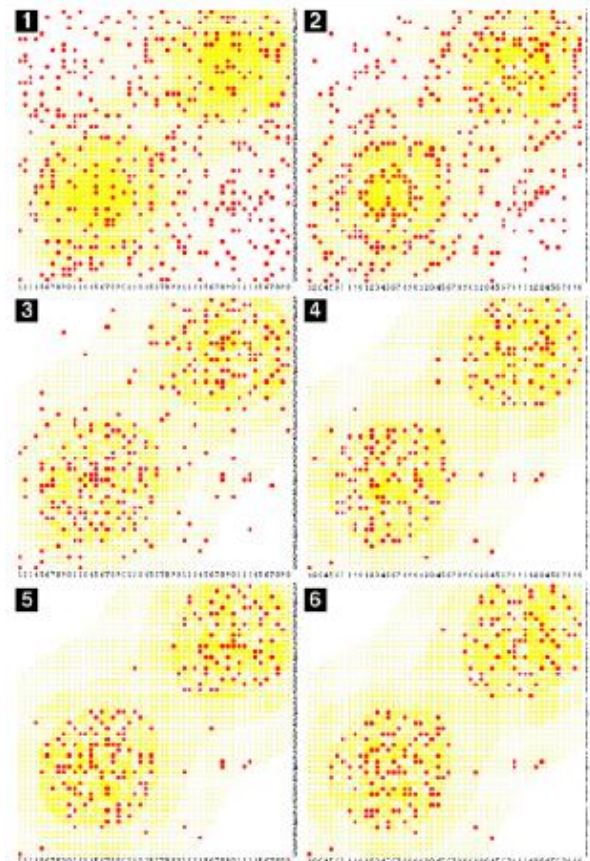

Fig. 3: Through simulation, agents eventually move to the areas of high sugar concentrations. Adapted from [3].

## 2 The Sugarscape Model

Writing an agent-based model begins with making assumptions on the interactions among the individuals or agents involved in the model. The agents are then simulated in an environment, producing and affecting global variables depending on their interactions and time. Tesfatsion [22] has discussed the specific goals that agent-based modelling researchers pursue,

- Empirical. Why particular large-scale phenomenon have performs like the emergence of social norms in society? Social norms patterns are analysed to see if they are replicated with various global irregularities to find out why behaviours persist.
- Normative. How can agent-based models be used to discover good designs? Key issues like efficiency and order are investigated.
- Heuristic. How can complex behaviours be attained through simple interactions? It is difficult to predict the behaviour of large-scale systems which are determined by the interactions taking place on the smaller scale.
- Methodological Advancement. Methods used by agent-based modelling researchers to study their models and simulation data.

The simple Sugarscape model allowed different scenarios to be modelled by adding simple extensions to include complex behaviour in society making it an ideal model to replicate how real societies work [13], [16]:
- Measuring wealth distribution. The sugars collected by the agents could be assessed to see what portion of the society was able to capture the most sugar.
- Disease propagation. Diseased agents are introduced into the scenario. During the simulation disease could be seen spreading across the landscape.
- Forming of social networks. Social networks could be formed between neighbouring agents or agents who collided into each other.
- Migration among the agents. Territorial areas could be established and agents could be seen moving across these territories similar to how migration works in the real world where people move in search for jobs.
- Sexual reproduction. The agents were given certain genetic material. Each agent would then scan its neighbours and choose a neighbour whose genetic material would be most similar to itself. Fertile agents were quickly seen to find each other and mate to produce new agents sprouting across the landscape.
- Inheritance among family members. With the concept of families, inheritance of wealth or sugar was programmed. This allowed various society classes to be formed.
- Combat. Sometimes agents were allowed to combat each other for commodities.
- Life and death. Agents were given life span to allow a living society to be modelled.
- Trading between sugar and spice. Wealth in the Sugarscape model was denoted by the amount of sugar. Spice was an additional commodity which was introduced by [8]. The agents were told they need specific proportions of both sugar and spice to survive. During the simulation if an agent had too much of sugar and *bumped* into another agent who had extra spice, the two agents would agree to *trade*. Through this model the supply and demand curves were generated, which showed that this proof can be used to analyse economic models.
- Terrains. Some models had mountains and landscape design which showed agents concentrating in areas of high sugar and which were easily accessible.

Figure 3 displays the movement of the agents across the landscape in a simple Sugarscape model. The agents concentrated into the two areas where there were piles of sugar in the beginning.

# 3 FLAME Model Implementation

## 3.1 FLAME Framework version 1.0
FLAME (Flexible Large-scale Agent-based Modelling Environment) is a tool which allows modellers from all disciplines, economics, biology or social sciences to easily write their own agent-based models. The environment is a first of its kind which allows simulations of large concentrations of agents to be run on parallel computers without any hindrance to the modellers themselves. This study uses the open source version 1.0 of the platform.

The FLAME framework is a tool which enables creation of agent-based models that can be run on high performance computers (HPCs) [5]. The framework is based on the logical communicating extended finite state machine theory (X-machine) which gives the agents more power to enable writing of complex models for large complex systems [6]. The agents are modelled as communicating X-machines allowing them to communicate through messages being sent to each other as specified by the modeller. This information is automatically read by the FLAME framework and generates a simulation program which enables these models to be parallelised efficiently over parallel computers.

The simulation program generator for FLAME is called the Xparser. The Xparser is a series of compilation files which can be compiled with the modeller's files to produce a simulation package for running the simulations. The files highlighted are as follows,
- Model.xml - Multiple xml files contain the whole structure of the model such as agent descriptions, memory variables, functions, messages.
- Functions.c - Multiple '.c' files contain the implementations of the agent functions specified in the xml files.
- 0.xml - This contains the initial states of the memory variables of the agents such as the initialisation of all parameters. The number of the resulting XML files depends on the number of iterations specified to run a model (through Main.exe).

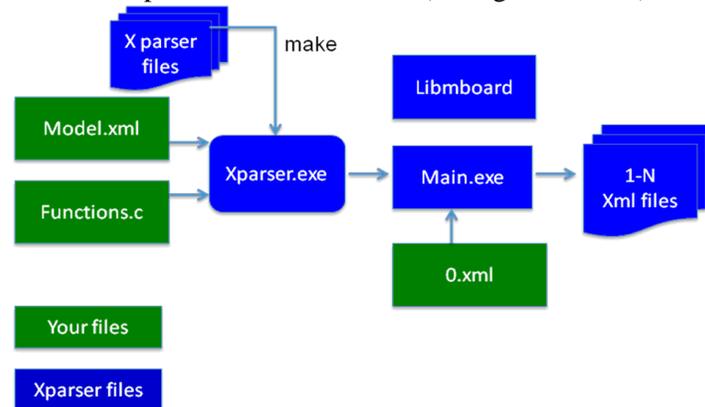

Fig. 4: Block diagram of the Xparser, the FLAME simulation program. Blocks in blue are the files automatically generated. The green blocks are modeler files.

### 3.2 X-machine agents

FLAME uses X-machines [14] to represent all the agents acting in the system. Each would thus possess the following characteristics:
- A finite set of internal states of the agent.
- Set of transitions functions that operate between the states.
- An internal memory set of the agent.
- A language for sending and receiving messages among agents.

Figure 5 shows the structure of how two X-machines will communicate. The machines communicate through a common message board, to which they post and read from their messages. Using conventional state machines to describe the state-dependent behaviour of a system by outlining the inputs to the system, but this failed to include the effect of messages being read and the changes in the memory values of the machine. X-Machines are an extension to conventional state machines that include the manipulation of memory as part of the system behaviour, and thus are a suitable way to specify agents.

Describing a system in FLAME includes the following stages:
- Identifying the agents and their functions.
- Identify the states which impose some order of function execution within the agent.
- Identify the input messages and output messages of each function (including possible filters on inputs).
- Identify the memory as the set of variables that are accessed by functions (including possible conditions on variables for the functions to occur).

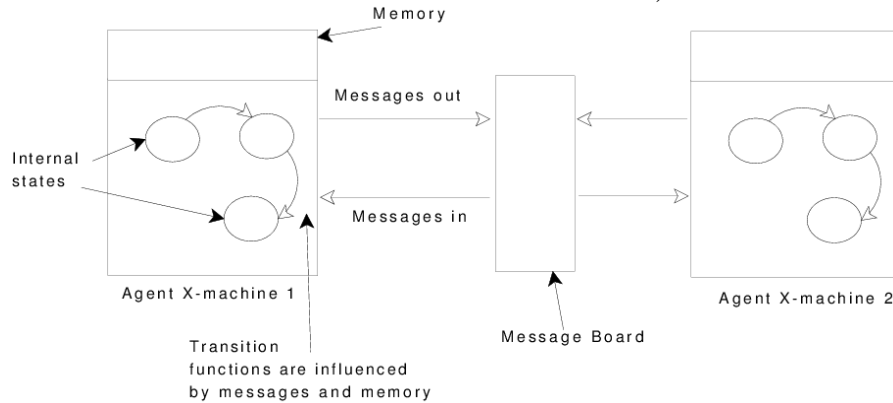

Fig. 5: How two agent x-machines communicate? The agents send and read messages from the message board which maintains a database of all the messages sent by the agents.

## 3.3 Writing the Model

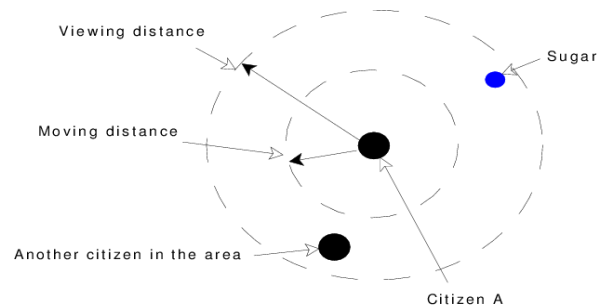

Fig. 6: View of a Citizen agent in FLAME Sugarscape.

In the Sugarscape model, the position of the agent and the sugars holds great influence on how interactions take place. Each agent or citizen has a certain area of vision and moving distance for capturing the sugar. This way most of the decisions are made locally depending on what happens in this area of influence of the agent. Figure 6 shows the area of influence of one agent (Citizen A) where agent decisions are based on local stimuli. The distance of moving can be set as one of the parameters of the model. Table 1 summarises the parameters used in the model.

Table 1: Model Global parameters in FLAME Sugarscape simulation.

| Parameter | (in units) |
| --- | --- |
| Viewing distance | 200 |
| Eating distance | 5 |
| Moving or run distance | 5.5 |
| Landscape | 200 x 200 |

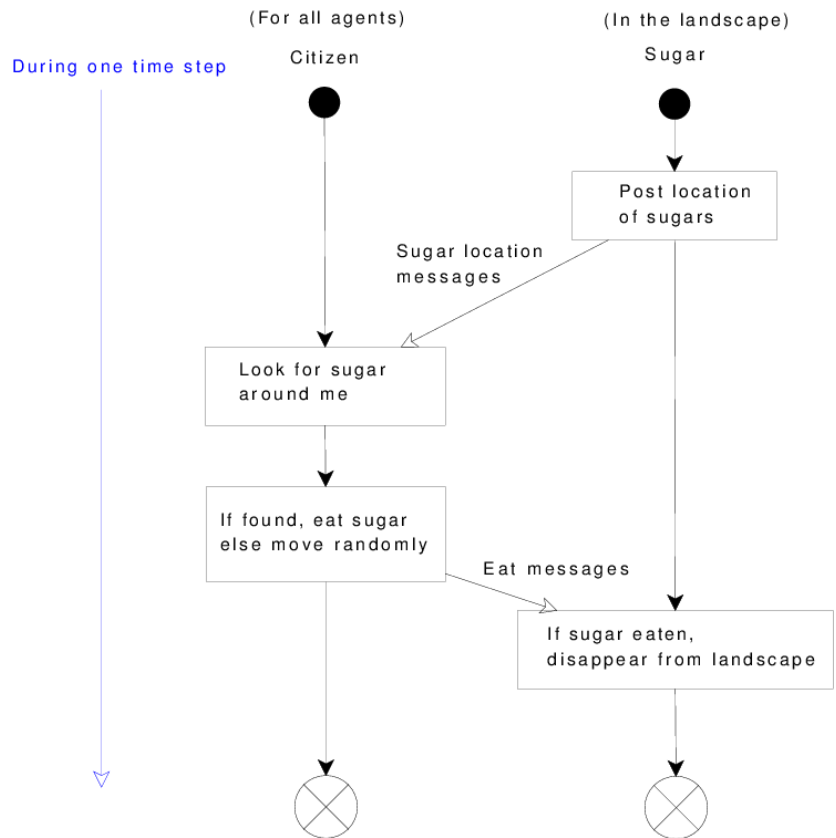

Fig. 7: Timeline of the basic FLAME Sugarscape model.

Figure 7 shows the basic functions of the agents performed in one iteration of the simulation. The sugars will post their locations globally, allowing the citizens to find them. If the sugar is eaten by the citizens, these disappear from the landscape. Note that the iteration timeline shown in the figure, does not consider the citizen agents dying or trading their sugar as observed in some of the literature review [7]. The same experiment was simulated with three different starting scenarios to measure if the wealth distribution varied among the agents. The 0.xml file was generated three times with three different settings - Random mixed distribution, Separate areas and Overlapping areas as shown in Figure 8. The model was then simulated with different settings for initial distributions about 20 times and then averaged to find the data patterns in the experiments. Tables 3, 4 discuss the algorithms for the agents which involves them posting and reading messages from each other.

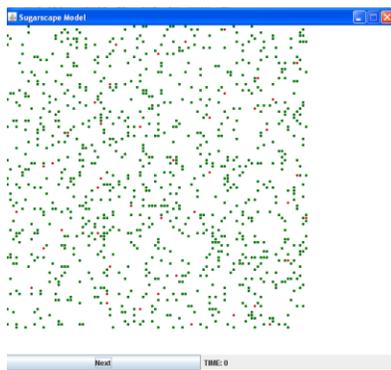
(a) Random mixed distribution of citizens and sugar.

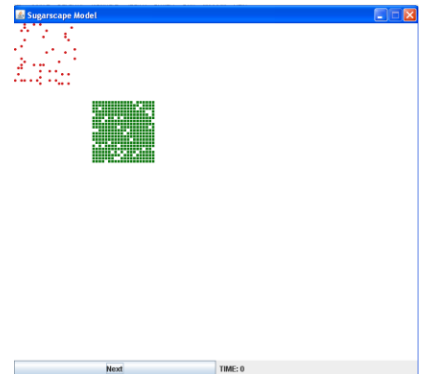
(b) Separate areas of citizens and sugars.

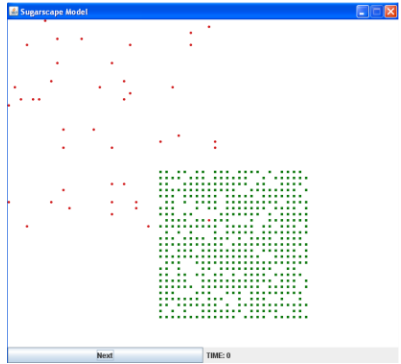
(c) Overlapping areas of sugar and citizens.

Fig. 8: Three different initial setting for the simple Sugarscape experiment. The citizen agents are represented by the *red* dots and the *green* dots represent the sugar agents in the scenario

### 3.4 Measuring Wealth Distribution in FLAME Sugarscape

Table 2 display the memory variables held by the corresponding agents. Although the sugar is an inactive entity, it needs to post its location so that it can be found by the other agents. This is because FLAME does not have an environmental agent which would possess this information for the agents to access it, they had to post their locations to the other agents. The sugar agents will perform the following functions:
- Post their location messages so that they are read by the citizen agents.
- Read 'eat' messages to determine if they have been eaten and should disappear (or die) from the landscape.

The citizen agent will also perform two functions as follows:
- Read the sugar messages to locate where they are.
- If the distance is close within the citizen's moving range, 'eat' the sugar.
- Send eaten messages out to sugar agents, so that they can 'die'.

Table 2: Memory of Citizen and Sugar agents.

| Citizen Agent | | Sugar Agent | |
|---|---|---|---|
| Type | Variable | Type | Variable |
| Integer | Id | Integer | Id |
| Integer | Sugars | Double | x |
| Double | X | Double | y |
| Double | y | Integer | Experiment id |
| Integer | Sugars collected | | |
| Integer | Flag if sugar collected | | |
| Integer | Experiment id | | |

The experiment was simulated with three different initial distributions to find whether just varying the initial distribution had an effect on the simulation results. Figure 8a represents a random distribution of citizens and sugars similar to the distribution used by Beinhocker [3]. Figure 8b represents two separate areas, one for sugars and one for the citizens lying far away from the sugar piles. And finally, Figure 8c displays an overlapping scenario where some agents lie close to the sugar while others lie far away. The three different setting shows the influence the distances to resources may have on the wealth of the populations arguing whether there is an additional dimension to resource availability to population wealth distributions when people have to travel to work. The graphs show the different amounts of sugar possessed by the agents at different times of the simulation. The graphs plotted, here, are logs of the frequency of citizens which had collected the different numbers of sugar[1].

Table 3: Algorithm for Sugar Agent.

> For all Sugar agents:
> For each iteration:
> Step 1: Post my location to the world. Post 'Location' messages to the environment.
> Step 2: Check if I have been eaten. Read 'Eaten' messages from Citizen agents.
> Step 3: If I am eaten die or disappear from the scenario.

Table 4: Algorithm for Citizen Agent.

> For all Citizen agents:
> For each iteration:
> Step 1: Look for Sugar agents. Read 'Location' messages from Sugar agents.
> Step 2: If sugar is found, move closer to Sugar to eat it.
> Step 2b: Else move randomly.
> Step 3: If Sugar is very close, then eat it. Post 'Eaten' messages to the environment.

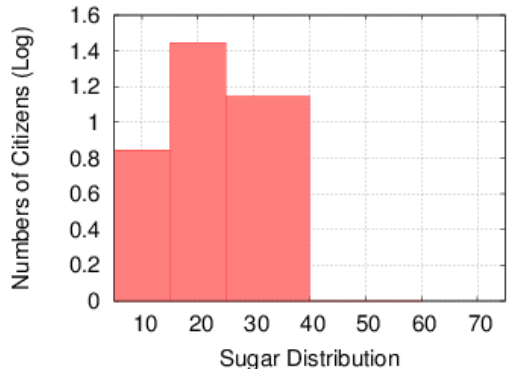
(a) At 100.

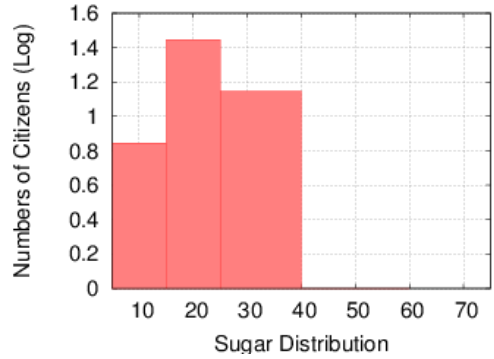
(b) At 200.

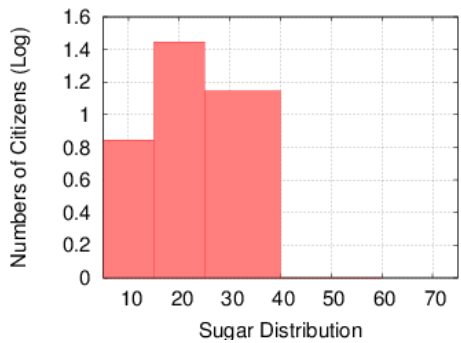
(c) At 350.

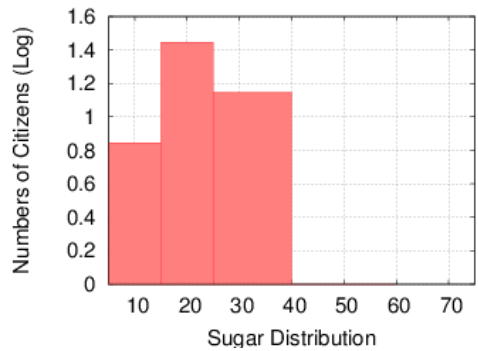
(d) At 500.

Fig. 9: Sugar collected for random initial distribution of citizens and sugars.

## 4 Results and Discussion

Beinhocker's results in Figure 15 depicted the Sugarscape as a fairly egalitarian society where the wealth distribution is a smooth bell-shaped curve. Starting with a small number of rich and poor agents, a broad middle class presented a small distance between the rich and the poor in the beginning. But with time few agents emerged as the super rich and the middle class shrink increasing the numbers of poor agents. Similar behaviour was not observed in these three experiments.

---

[1] A comparison with graphs for representing frequency versus log graphs has been presented in the end of the paper.

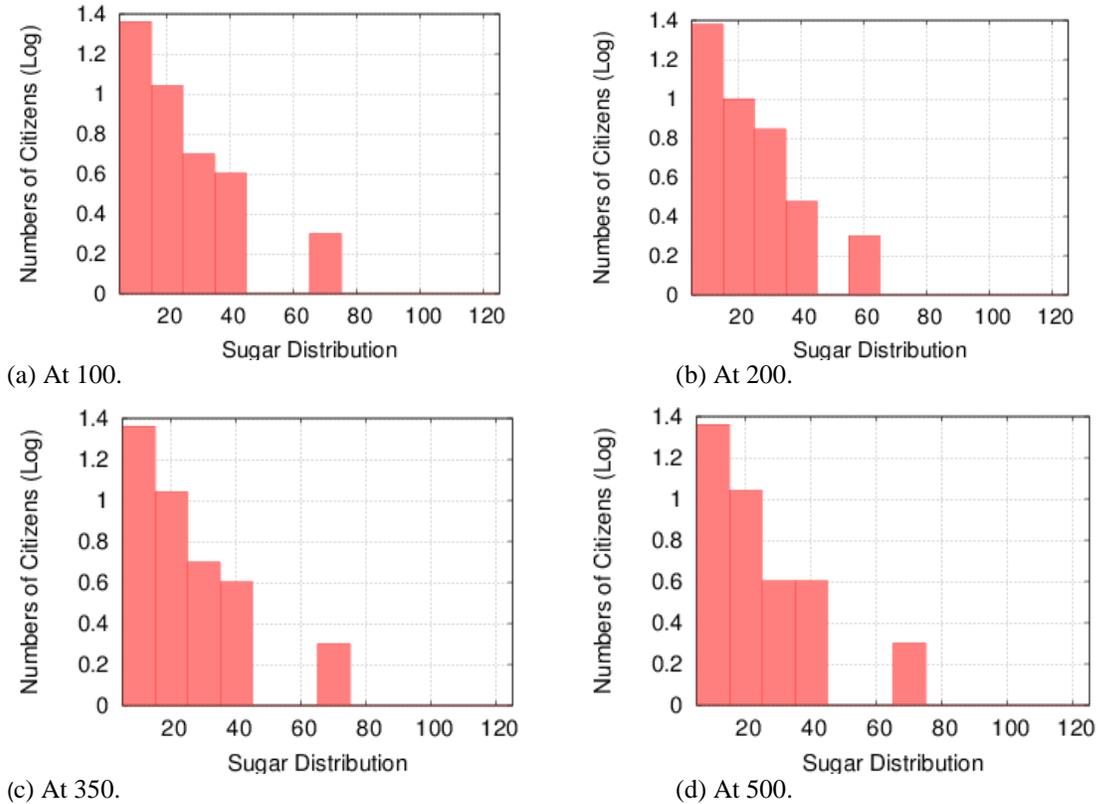

(a) At 100.  (b) At 200.
(c) At 350.  (d) At 500.

Fig. 10: Sugar collected for separate areas initial distribution of citizens and Sugars

Figures 9, 10 and 11 showed that the initial conditions influenced the wealth distribution among the citizen society. Figure 9 shows the agents were quick to grab the sugar around them displaying a positive skewness with virtually no poor agents. Figures 10 and 11 show a slower progress on capturing the sugar. These figures also show a positive skewness with a distribution of all kinds of class distribution with a large distribution between the rich and poor agents.

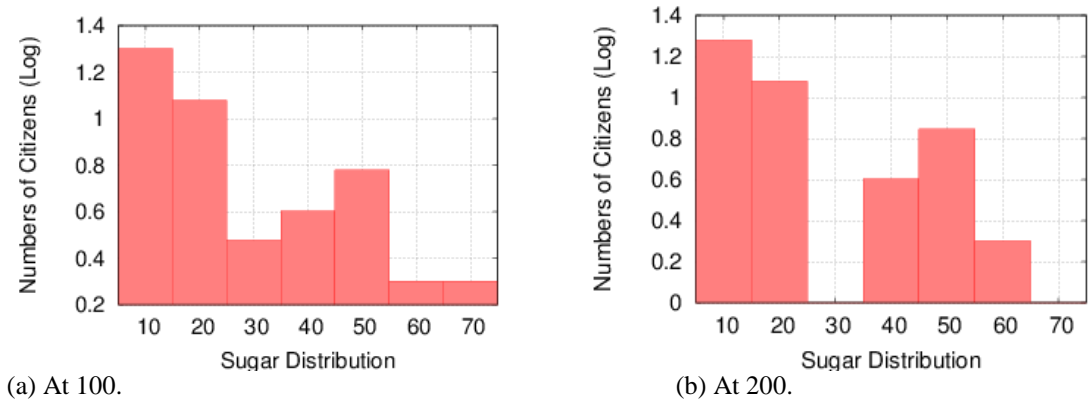

(a) At 100.  (b) At 200.

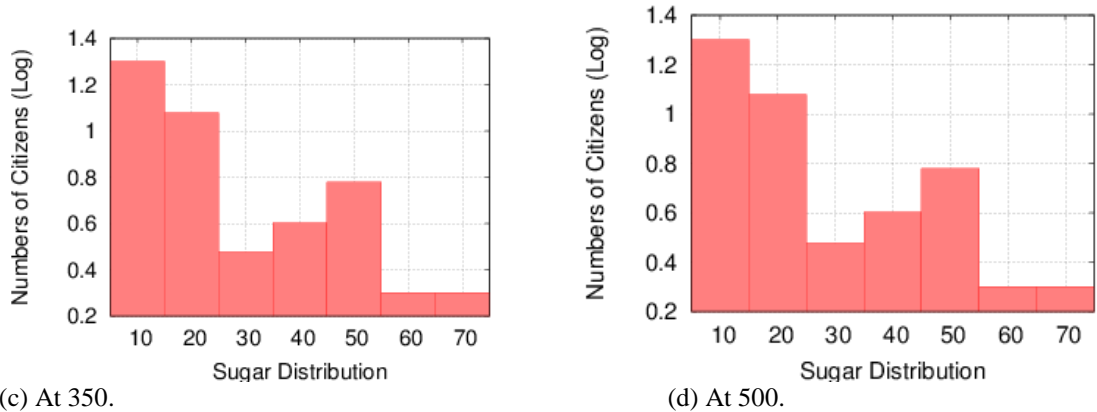

(c) At 350.  (d) At 500.

Fig. 11: Sugar collected for overlapping areas initial distribution of citizens and sugars.

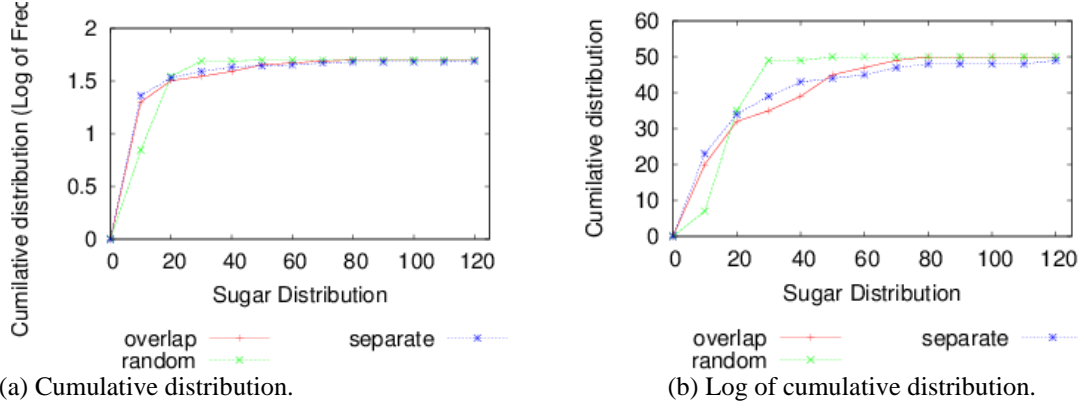

(a) Cumulative distribution.  (b) Log of cumulative distribution.

Fig. 12: Distribution of agent numbers with captured sugar.

The results were analysed for the skewness and kurtosis in the data. Skewness is the measure of asymmetry of the data distribution. Kurtosis is the measure of the bulge or peakness of the data distribution.

Table 5: Results of skewness and kurtosis measures of the three experiments.

| Initial distribution | Random | Separate Areas | Overlapping areas |
| --- | --- | --- | --- |
| Skewness | 1.586 | 2.418 | 1.530 |
| Kurtosis | 2.047 | 6.043 | 1.692 |

Table 5 summarises the skewness and kurtosis of the distribution of the sugar collected at time $t = 500$ of the simulation. The higher the kurtosis the greater the distribution between the rich and the poor. The experiment with separate areas of citizens and sugars displayed the highest difference between the rich and the poor in the society. This was followed by the random distribution results and lastly the overlapping area results.

Figure 12 displays the cumulative distribution of the sugars gathered during the simulation. The maximum amount of sugar held by any one of the 50 citizens was 120.

$$\frac{20}{120} \times 100 = 24$$

Figures 16, 17 and 18 display the frequency of the sugars gathered. The bar charts across the three graphs show that most of the agents were able to capture sugar in the range between 10-30, showing the 80% population wealth. Only Figure 16 with the random locations were able to capture sugars between 20-30 supported by the equation for 24 sugars. Figure 12b depicts a pattern similar displaying 80% of the wealth is held by 20% of the population. The plots for the three experiments show that most of the sugar was captured between the 20-30 agent distribution. However, the experiments for overlapping and separate areas took longer to achieve this.

## 4.1 Simulation times of the three Case studies

The experiment involved 21020 agents with 50 citizens, 1000 sugars and an Averager agent which monitored 20 scenes of the experiments. The time taken to run each scenario was measured and is displayed in Figure 13. The experiments were run in serial and in parallel on the Mac Laptop machine. The parallel distribution was done in both geometric partitioning (partitions on geographic distribution using x and y positions of agents) and round robin partitioning (partitions on agent numbers on available processors).

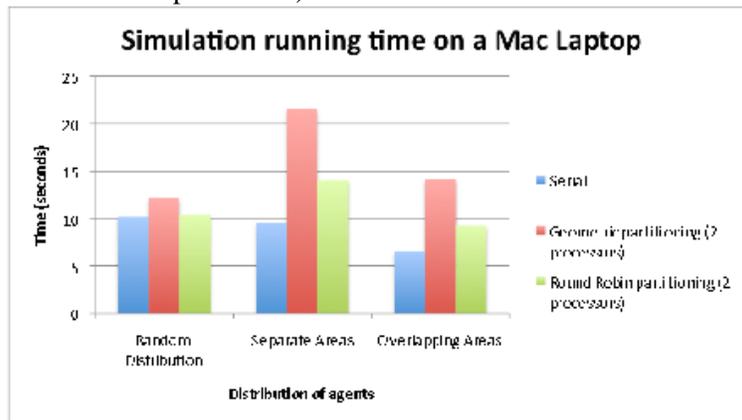

Fig. 13: Simulation running times in the Sugarscape model by only changing the initial conditions.

Figure 13 shows that even though the number of agents were the same, by changing the initial distribution, the simulation time changed. The random distribution took the least time because the agents, citizens and sugars could communicate locally on the same processors. However, when they were separated into different areas the agents had to communicate over different processors, thus causing an increase in the simulation time. A similar result was seen in the overlapping areas being more than the random distribution times. These results showed that the messaging between agents holds key for the simulation times in agent-based modelling approaches. How the agents are distributed influences if they have to communicate locally on the same processor or across nodes which increases the simulation time for messages to be sent across.

## 5 Related Work

The Sugarscape model proved to be an excellent tool to analyse economic models in an artificial society of agents. Horres et al. [12] discussed the similarities between economics in real and artificial societies using the spice trading facet of the model. The authors presented various results of the supply demand curve depicting how the equilibrium changes with different experimental settings. Gumerman et al. [10] used the Sugarscape model to produce results which were later mapped to show how the prehistoric American society settlements were organised in history. Klock [15] presented a detailed report on the implementation of his version of the Sugarscape model which was based on the formation of territories. His work investigated territorial behaviour of trade, combat and wealth distribution when various factors like migration and taxes were introduced. Al et al. [1] used the model to study the effects of taxing wealth and redistribution when they measured the collected taxes over the population of 400 agents, showing that, using high tax rates was good for the population to survive but poor agents struggled. Similar results was shown by [2] and [1]. Buzing et al. [4] showed how learning and communication could influence the agents where only a certain population of agents were allowed to learn new strategies from other agents. Their results showed that evolution only influenced those agents who listened. Increasing the communication among the agents increased cooperation among the population depicting that societies with no or little communication find it difficult to survive. These results are similar to those provided by [17]. Hales [11] used the Sugarscape society with memetic algorithms to display the propagation of cultural information among the population of agents.

Sugarscape's key advantages lie in the phenomenon that the model could be tweaked with various variables or methods to allow new behaviours to be introduced in the agents. By introducing authorities or leaders governments could be seen emerging. Peterson [18] supported the use of the model saying, "by providing insights into population growth, resource use, migration, economic development, conflict, and other global social processes, games played on the Sugarscape grid may help shape the policies needed to direct the future course of society".

The Sugarscape was also heavily criticised in [19] saying that it was too restrictive to be used as an economic analysis model. This is because some of the results of the model showed the lack of steady state behaviour in certain scenarios. They claimed that as Sugarscape omitted existing economic theories it could not be used for testing. Beinhocker [3] argued that [7] had not expected that Sugarscape would become a model for economics. Yet, it was able to produce striking results free of the unrealistic assumptions found in traditional economics. It is not being based on the equilibrium system and neither does it go into it. It is a useful model which displays complex structures evolving from bottom up from simple starting rules at low level interactions. Scientists have experimented with a number of other example models to mimic some aspects of real societies such as Conway's Game of Life, as one of the earliest examples using cellular automata grids to display a group of cells interacting with each other [9]. Using only four simple rules through simulations, the game was able to present new emerging patterns depending on the neighbouring cells. The game loops through the cells and check for these four rules: (1) If current cell is alive and it has less than two neighbouring cells, die due to lack of social activity. (2) If current cell is alive and it has more than three neighbouring cells, die due to over crowding. (3) If current cell has 2 or 3 live neighbours around, survive to the next time step. (4) If current cell is a dead cell but has three live neighbours, become alive.

Another example is the segregation model to demonstrate how, by adding a small preference factor, societies can emerge into following segregated patterns [20]. The model was based on a 2D grid landscape with individuals represented by different colours where every coloured individual would check for the rule, which if more than 33% of its adjacent individuals or cells are of a different colour the cell would randomly move to a new position. This model was a pioneer in social science as it demonstrated that by just adding a small factor of 33% for neighbour preferences, the societies would eventually separate displaying how real societies emerge. However if this preference factor was increased to 50%, the model would not show this behaviour. Individuals would then have a 50-50 preference for their neighbours allowing societies to accept their neighbours giving them an equal opportunity to consider staying.

## 5.1 Evolution from Bottom-up

Sugarscape is also a useful model to see how societies develop and evolve. If the agents were equipped with genetic material this could be inherited by a new born child agent, allowing the best genes to be carried on to new generations. It has been shown that when running such an experiment, 'over a period of time, the characteristics of the population of agents converge towards certain traits, namely good vision and a low metabolism' [16].

Similar institutions like banks show economic trading as an example of emerging evolution. This was observed in Sugarscape when [8] introduced the following rules into the scenario:
- An agent could be a lender if it is too old to have children or if it has more savings than it needs for reproductive purposes.
- An agent could be a borrower if it has insufficient savings to produce children, but a sufficient supply of one of the two sugar or spice.
- An interest rate could be added on the loans which could then be collected when returned.
- If agents were credit worthy they were allowed to borrow sugar for their own needs.

When Epstein and Axtell, plotted the relationships between the lenders and borrowers they were able to trace complex relationships between the rich and the poor. The relationships showed that the rich agents were lending to poorer agents through middle agents who were performing functions similar to how banks behave in real life (Figure 14).

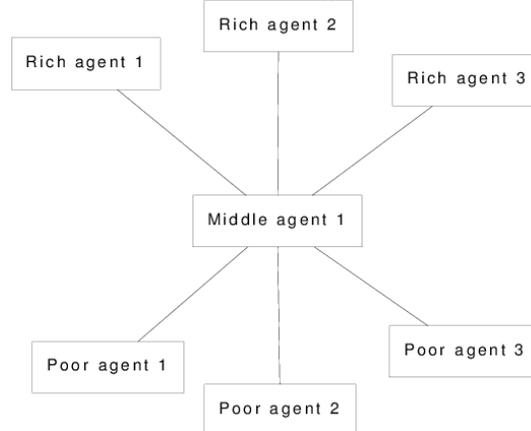

Fig. 14: Relationships emerged between the rich and the poor agents. The middle agents behaved similar to how banks behave in real life.

## 5.2 Distribution of Wealth

Traditional economic theories follow the Pareto laws which states that markets always lead to perfect allocation of resources among the population. Beinhocker [3] used the Sugarscape model to display this pattern in the sugar distribution across a set of agents with Figure 15 showing the cumulative distribution of the sugars. The figure supports the economic inequality among the agents displaying the right hand tail stretching to have only a few rich agents. The *bump* formed in the middle in the beginning of the simulation slowly shrinks as time progresses. However the model displays no relation between the cause and effect as to why some agents are poorer than others and what could be the reasons for this inequality.

Being an emergent property from a given distribution, these results support the invisible hand phenomenon of economics. However it is possible that the initial distributions may have an effect on how agents become wealthy. Agents lying closer to sugar areas would become richer quicker than agents situated far away.

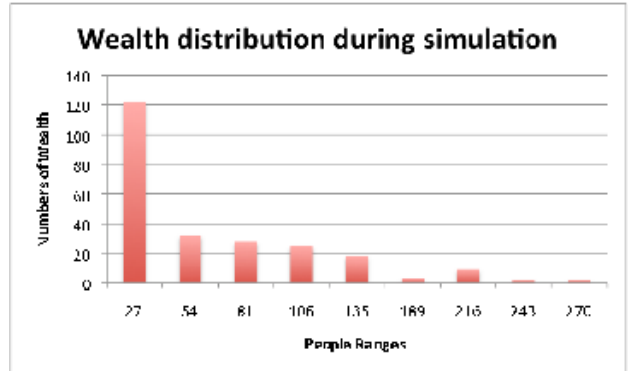

Fig. 15: Through simulation wealth distribution among the agents for a randomly initial distribution of agents. c.f. [3].

# 6 Conclusions

This paper was inspired by the Sugarscape model, such that the same model was simulated three times, with different initial settings, varying the distances of the sugars to the agents. This was done to test whether the *initial sugar grabbing* would affect the overall results of the wealth distribution and have an influence on the simulation times of the model as well. The three results obtained showed different patterns showing that distances were able to affect the wealth distributions among the citizens, creating larger middle classes in the scenarios. The random distributions were able to grab sugar more quickly with less poor agents, in comparison to the other two cases where slower to capture sugar with larger poor and middle classes. The results show how resources affect wealth among populations.

The experiment could be enhanced in the future by adding more factors like metabolism rates and further facets collected from literature to build a complete artificial society, to test further economic principles like labour markets or governments affecting how citizens behave in the system, to mimic real human societies.

The experiment also showed that despite writing the same model, just varying the initial distribution of the agents over processors can largely influence the simulation times of the models. This shows the communication overhead play a key role when parallelisation algorithms are designed and when writing models to optimise this, frameworks should consider the initial distributions of the model specification (i.e. agents communication) as well with efficient parallelisation of the model itself.

Whether Sugarscape is a useful model for testing economic models is a debatable issue. However it is a good starting point for modelling economic activities where location of the agents would highly influence the agent behaviour.

Most economic models do not have the concept of location to see whether this affects the distribution among the agents.

# Appendix

```
#include "header.h"
#include "my_library_header.h"
#include "Sugar_agent_header.h"

int Sugar_post_location()
{
```

```
        add_sugar_location_message(ID, X,Y, SCENE_ID);
        return 0;
}

int Sugar_check_eaten()
{
        int citizen_id=-1;
        request_sugar_message=get_first_request_sugar_message();
        while(request_sugar_message)
        {
            if(request_sugar_message->scene_id==SCENE_ID)
            {
                if(request_sugar_message->sugar_id==ID)
                {
                    citizen_id=request_sugar_message->citizen_id;
                }
            }
            request_sugar_message=get_next_request_sugar_message(request_sugar_message);
        }
        if(citizen_id!=-1)
        {
            add_eaten_message(citizen_id, X, Y,SCENE_ID);
            return 1; // kill off agent
        }
        return 0;
}
```

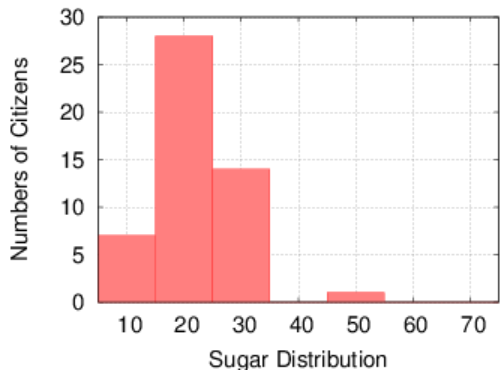
(a) At 100.

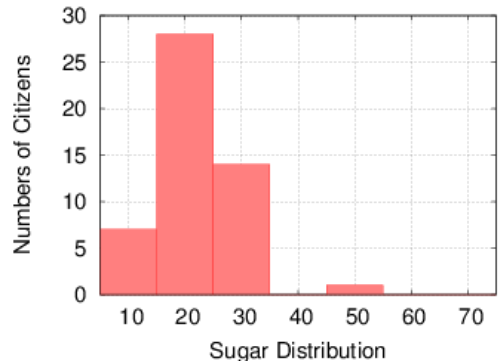
(b) At 200.

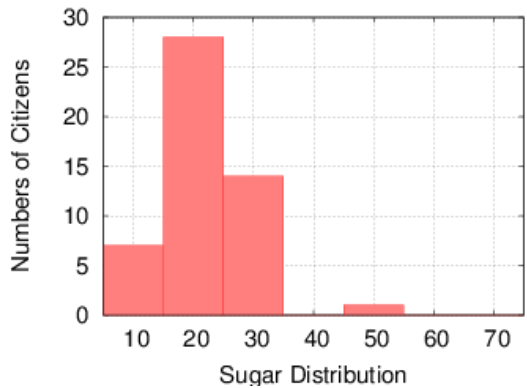
(c) At 350.

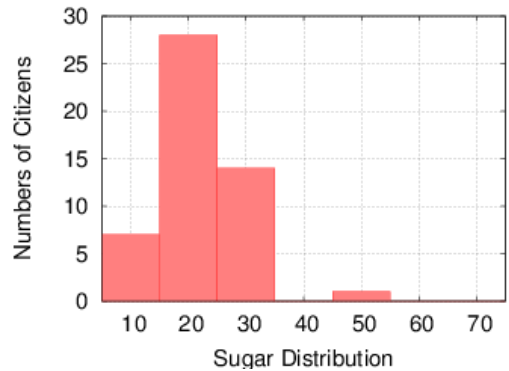
(d) At 500.

Fig. 16: Frequency sugar distribution for random distribution of citizens and sugars.

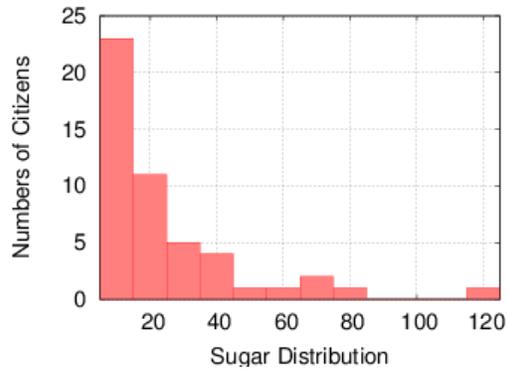
(a) At 100.

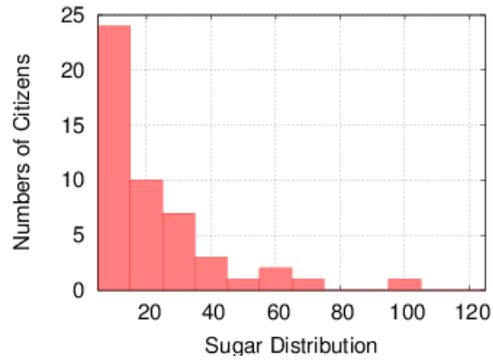
(b) At 200.

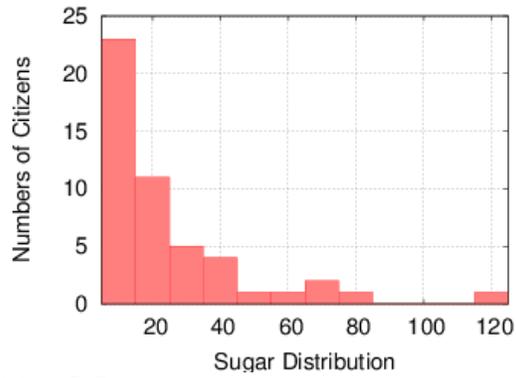
(c) At 350.

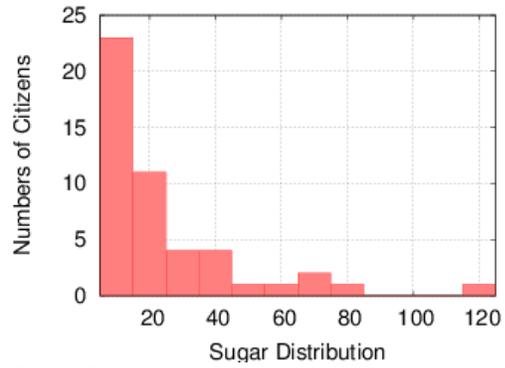
(d) At 500.

Fig. 17: Frequency sugar distribution for separate distribution of citizens and sugars.

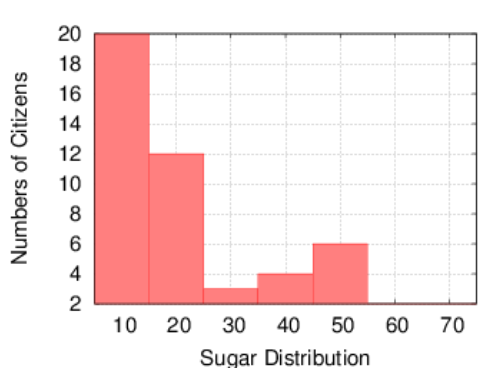
(a) At 100.

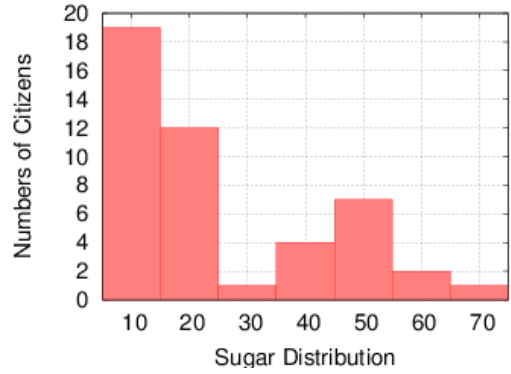
(b) At 200.

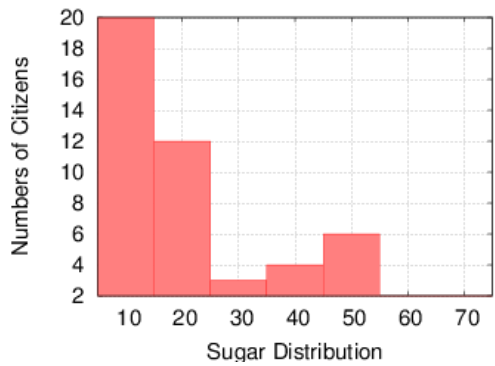
(c) At 350.

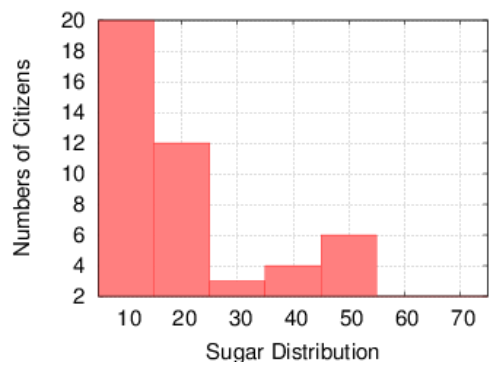
(d) At 500.

Fig. 18: Frequency sugar distribution for overlap distribution of citizens and sugars.